\documentclass[letterpaper]{article}
\usepackage{spconf,amsmath,amssymb,graphicx,booktabs,tabularx,hyperref,capt-of,balance,cite}
\usepackage{flafter}
\usepackage{iftex}
\RequireXeTeX
\usepackage[no-math]{fontspec}
\newfontfamily\cjkfont[Path=./]{FandolSong-Regular.otf}
\usepackage[T1]{fontenc}

\hypersetup{hidelinks,pdftitle={HMU-Bench: Hierarchical Music Understanding Benchmark for Omni Models}}
\title{HMU-Bench: Hierarchical Music Understanding\\Benchmark for Omni Models}
\name{\shortstack{\href{https://orcid.org/0009-0002-4780-4666}{Zheqi Dai}$^{1,*}$ \quad
\href{https://orcid.org/0009-0008-4399-534X}{Boyang Wang}$^{2,*}$ \quad
\href{https://orcid.org/0009-0003-4765-2049}{Zhiyue Wu}$^{3}$ \quad
\href{https://orcid.org/0009-0000-6151-3465}{Yiyang Xu}$^{4}$ \quad
\href{https://orcid.org/0009-0009-4979-9611}{Xinming Shi}$^{5}$\\
\href{https://orcid.org/0009-0002-0539-6916}{Ruibin Yuan}$^{6}$ \quad
\href{https://orcid.org/0009-0007-4720-1968}{Xuerui Yang}$^{3}$ \quad
\href{https://orcid.org/0009-0005-2897-0011}{Chao Yan}$^{3,\dagger}$ \quad
\href{https://orcid.org/0000-0003-2864-0475}{Qiuqiang Kong}$^{1,\dagger}$}%
\thanks{\small $^{*}$Equal contribution. $^{\dagger}$Co-corresponding authors.}}
\address{$^{1}$The Chinese University of Hong Kong \qquad
$^{2}$University of California San Diego \qquad $^{3}$StepFun\\
$^{4}$Tsinghua University \qquad
$^{5}$Shanghai Conservatory of Music \qquad $^{6}$MAP}
\hypersetup{pdfauthor={Zheqi Dai, Boyang Wang, Zhiyue Wu, Yiyang Xu, Xinming Shi, Ruibin Yuan, Xuerui Yang, Chao Yan, Qiuqiang Kong}}

\begin{document}
\ninept
\raggedbottom
\maketitle
\begin{abstract}
Music-understanding benchmarks should identify both the evidence available to a model and the operations a correct answer requires. We introduce HMU-Bench, a bilingual benchmark with notation and audio components. ABC Q\&A contains 1,029 Chinese--English question pairs across six musical domains, labeled as extraction, single-step inference, or multi-step inference. Audio Q\&A contains 460 question pairs on 92 recordings, spanning five scope--difficulty configurations. All music is Suno-generated, and questions and references are human-reviewed. Reference-based evaluation of seven omni models yields high notation scores for the strongest systems, but no model averages above 1.4 on the 0--4 audio scale. Notation and audio rankings show little correlation. A transposition case study also reveals octave errors missed by LLM grading. These findings motivate separate modality reporting and event-level checks for symbolic outputs. Data, annotations, evaluator prompts, and graded responses will be released at \url{https://github.com/stepaudiollm/hmu-bench}.
\end{abstract}

\section{Introduction}

Music understanding means reading musical elements and the relations that link them over time. Whether a model is used for analysis, teaching, or creative assistance, it must identify notes, rhythms, and chords, explain what they do, and connect local detail to the whole piece. A benchmark should therefore state both the evidence a model receives and the operations a correct answer requires. This matters especially for omni models, which accept audio as well as text.

Existing benchmarks cover parts of this problem. On the symbolic side, MusicTheoryBench, released with ChatMusician~\cite{chatmusician}, tests theory and reasoning through multiple-choice questions, including ABC-based reasoning~\cite{walshaw2011abc}. ABC-Eval evaluates symbolic understanding and instruction following~\cite{abceval}; SSMR-Bench synthesizes sheet-music problems with rule-verifiable answers~\cite{ssmr}; and CSyMR-Bench tests compositions of atomic score analyses~\cite{csymr}. MSU-Bench uses generative questions on complete classical scores in ABC and PDF at four levels~\cite{msu}. These resources establish complementary tests of notation understanding.

On the audio side, MuChoMusic uses multiple-choice questions over public captioned music~\cite{muchomusic}; a follow-up identifies questions answerable without listening and filters for perceptual dependence~\cite{rulistening}. MUSE probes perception and relational reasoning~\cite{muse}, BASS targets structure and semantic reasoning~\cite{bass}, MusicBench evaluates global and temporal understanding~\cite{gamma}, and MusTBench and MusicGroundingBench focus on temporal grounding~\cite{mustbench,musicgrounding}. MusICA-MetaBench extends multiple-choice perception tests across musical representations~\cite{musica}. HMU-Bench complements these efforts with open-ended, human-reviewed references and per-question solution labels, evaluating the same omni models on notation and audio. The labels help identify \emph{where} models fail, beyond an overall score.

We introduce \textbf{HMU-Bench}, which evaluates omni models on notation and on audio under one scoring protocol. \textbf{ABC Q\&A} contains 1,029 Chinese--English question pairs over identical ABC excerpts, covering six musical domains and 38 task types; each question carries a solution-requirement label (extraction, single-step inference, or multi-step inference) stating what a correct answer must do. \textbf{Audio Q\&A} contains 460 Chinese--English question pairs on 92 recordings and crosses global/local scope with easy/hard and easy/medium/hard requirements, respectively. Answers are open-ended and may include an explanation or rewritten notation (Table~\ref{tab:example-item}). We use Suno-generated material rather than selecting recordings from public music corpora; recordings carry human chord and section annotations. ABC questions and references are model-drafted, while audio questions instantiate human-written templates and take their references from the annotations; all items are human-reviewed against their evidence. Our contributions are:

\textbf{(1) Benchmark.} A two-part, hierarchically labeled music-understanding benchmark: bilingual ABC questions labeled by solution requirement, and bilingual audio questions labeled by scope and difficulty, with references grounded in human annotations.

\textbf{(2) Construction pipeline.} An evidence-linked workflow from generated music to reviewed Q\&A: notation normalization, human chord and section annotation, chord-corrected transcription, model drafting for ABC, template instantiation for audio, and human review.

\textbf{(3) Evaluation.} Results for seven omni models and two text-only references under shared reference-based scoring: high notation performance for the best models, uniformly low audio performance, little correlation between notation and audio rankings, and cases in which LLM scoring misses octave errors in symbolic output.

\begin{figure*}[t]
\centering\includegraphics[width=\textwidth]{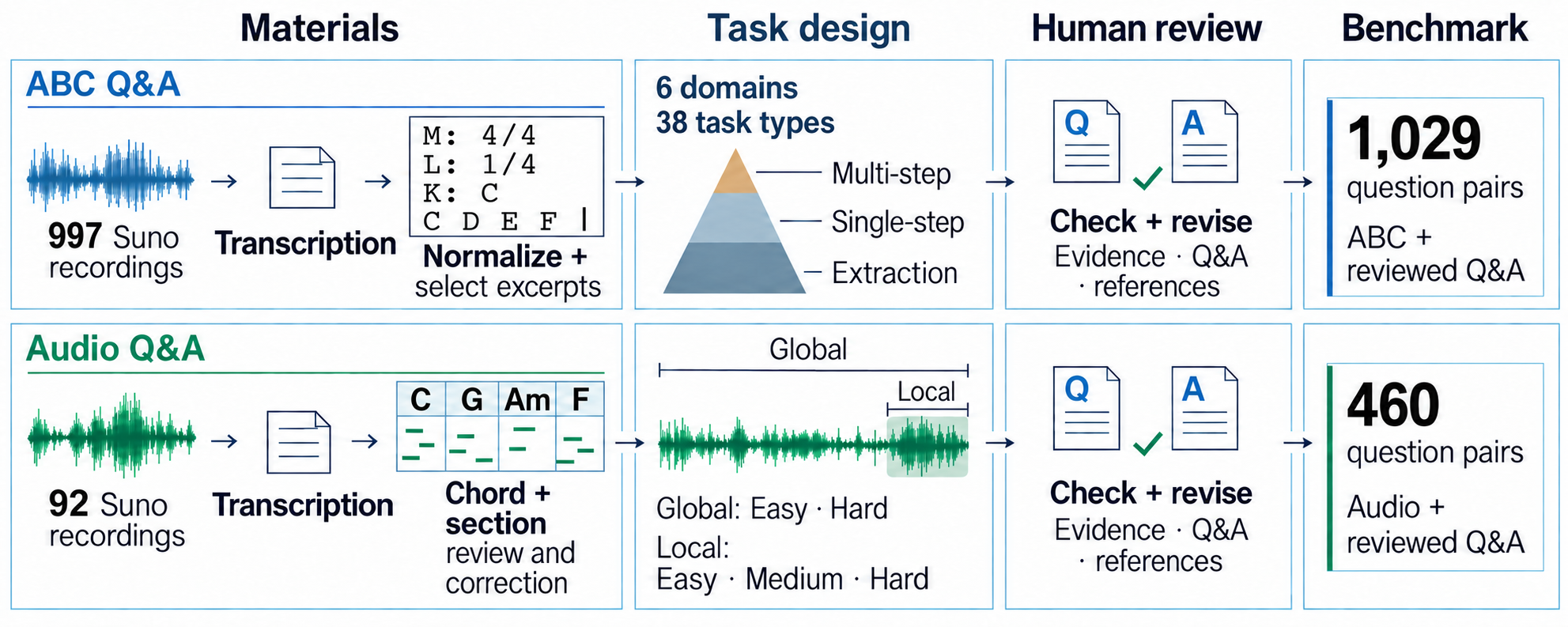}
\caption{HMU-Bench construction. ABC Q\&A uses normalized SheetSage2 excerpts; Audio Q\&A uses transcriptions refined with human chord and section annotations. Both paths undergo human review.}\label{fig:construction}
\end{figure*}

\begin{figure*}[t]
\centering\includegraphics[width=\textwidth,trim=8bp 112bp 8bp 122bp,clip]{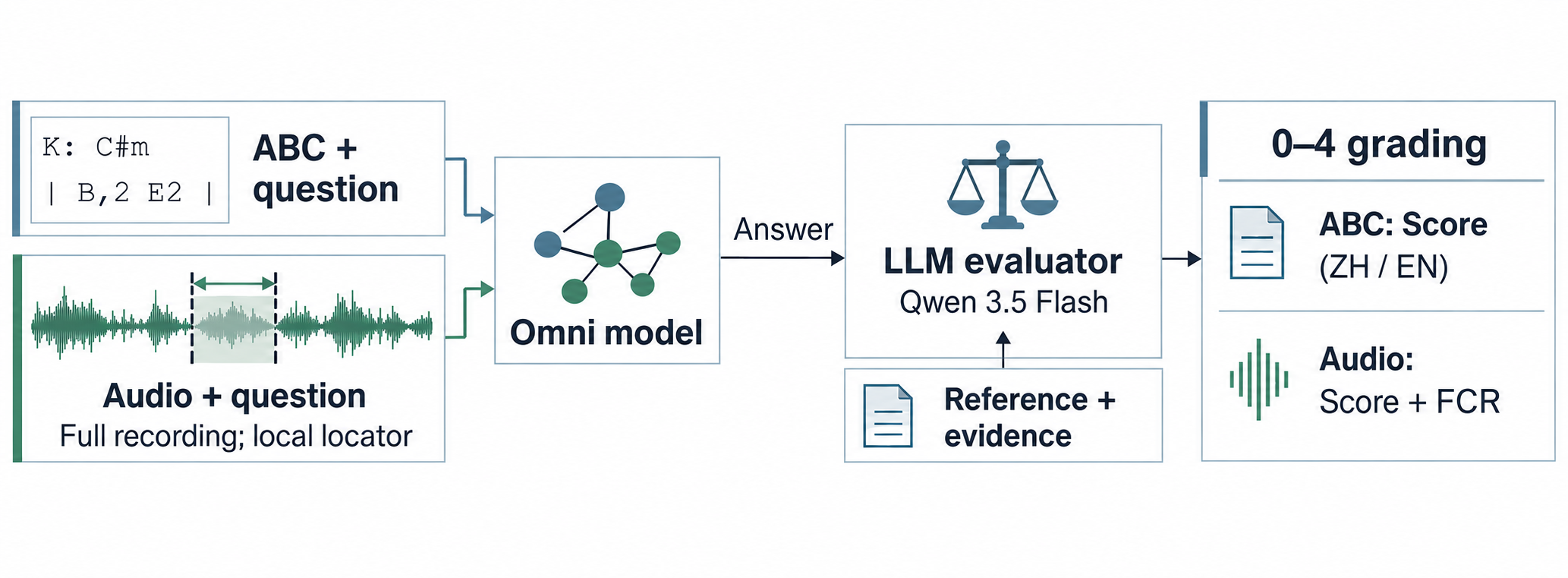}
\caption{Evaluation workflow. References and annotations enter only the grader; notation and audio scores are reported separately.}\label{fig:usage}
\end{figure*}

\begin{table*}[!t]
\centering
\caption{ABC Q\&A scores. ZH: Chinese; EN: English. Bold marks the best omni model; subgroup scores average both languages.}
\label{tab:abc-results}
\begingroup\fontsize{9}{10.5}\selectfont
\setlength{\tabcolsep}{2.3pt}\renewcommand{\arraystretch}{1}
\begin{tabular*}{\textwidth}{@{\extracolsep{\fill}}lrrrrrrrrrrr@{}}
\toprule
Model & \multicolumn{2}{c}{Overall score} & \multicolumn{6}{c}{Musical domain} & \multicolumn{3}{c}{Solution requirement}\\
\cmidrule(lr){2-3}\cmidrule(lr){4-9}\cmidrule(l){10-12}
 & ZH & EN & Global & Pitch & Harmony & Rhythm & Form & Symbolic & Extract. & Single & Multi\\
 & (1,029) & (1,029) & (49) & (268) & (392) & (208) & (25) & (87) & (335) & (441) & (253)\\
\midrule
Gemini 3.7 Flash & \textbf{3.840} & \textbf{3.864} & \textbf{3.97} & \textbf{3.69} & \textbf{3.91} & \textbf{3.95} & \textbf{3.76} & \textbf{3.83} & \textbf{3.81} & \textbf{3.86} & \textbf{3.89}\\
Seed Lite & 3.516 & 3.521 & 3.79 & 3.34 & 3.52 & 3.72 & 3.40 & 3.47 & 3.56 & 3.51 & 3.48\\
StepAudio 3 & 3.306 & 3.408 & 3.64 & 3.12 & 3.38 & 3.57 & 3.38 & 3.32 & 3.38 & 3.35 & 3.35\\
Gemini 3.1 Pro & 3.274 & 3.236 & 3.73 & 2.95 & 3.45 & 3.24 & 3.16 & 3.12 & 3.11 & 3.26 & 3.44\\
Gemini 2.5 Pro & 2.912 & 2.979 & 3.45 & 2.54 & 3.04 & 3.16 & 3.10 & 2.93 & 2.78 & 2.96 & 3.14\\
Qwen 3.5 Omni & 2.740 & 2.621 & 3.33 & 2.40 & 2.72 & 2.81 & 2.94 & 2.60 & 2.50 & 2.77 & 2.77\\
Qwen 3 Omni & 1.914 & 1.898 & 2.47 & 1.72 & 1.84 & 2.18 & 2.46 & 1.64 & 1.73 & 2.00 & 1.98\\
\midrule
\multicolumn{12}{l}{\textit{ABC-only reference systems}}\\[1pt]
GPT-6 Astra & 3.964 & 3.945 & 3.99 & 3.93 & 3.96 & 3.98 & 3.84 & 3.93 & 3.95 & 3.95 & 3.96\\
DeepSeek V4.1 & 2.982 & 3.033 & 3.58 & 2.60 & 3.19 & 3.07 & 3.00 & 2.95 & 2.87 & 3.04 & 3.12\\
\bottomrule
\end{tabular*}
\par\vspace{3pt}\begin{minipage}{\textwidth}\fontsize{9}{10.5}\selectfont
Counts denote question pairs. Single/Multi denote single-step/multi-step inference. Names omit Plus (Qwen 3.5 Omni), Flash (Qwen 3 Omni, DeepSeek V4.1), and Realtime (StepAudio 3); Seed Lite is Seed2.0 Lite.
\end{minipage}\endgroup
\end{table*}

\section{HMU-Bench Construction}

\label{sec:construction}

Figure~\ref{fig:construction} summarizes both construction paths.

\subsection{Data Sources and Annotation}

The ABC pool contains 997 Suno-generated songs automatically transcribed with \href{https://huggingface.co/m-a-p/SheetSage2}{SheetSage2}~\cite{yuan2025yue,li2024mert}. We align melody events to timing and beats and normalize chord symbols, inline key changes, and ties. Task-specific excerpts preserve note spelling, register, durations, rests, meter, and required harmonic context; lyrics, descriptive section names, and free-form comments are removed to reduce irrelevant textual cues.

The audio set contains 92 Suno recordings. Annotators listening to each recording identify or correct chords and section boundaries, and the corrected chords are used to adjust the SheetSage2 ABC transcription. Annotations and adjusted notation serve as the evidence for instantiating questions and constructing references, and section boundaries define the windows for local questions. All songs were generated by the authors on paid Suno subscriptions and obtained through Suno's download channel; the benchmark is intended for evaluation-only use, subject to the applicable Suno terms~\cite{suno_tos}.

\subsection{Task Design}

ABC Q\&A covers 38 task types in six domains: global information, pitch and melody, tonality and harmony, rhythm, form and temporal relations, and verifiable symbolic tasks (e.g., key identification, interval calculation, chord-function analysis, rhythmic comparison, notation parsing, and transposition). Each question is labeled by what a correct answer must do. \textbf{Extraction (e)} reads a property encoded in the notation, such as the entry beat after a rest. \textbf{Single-step inference (1)} applies one musical relation: under \texttt{K:E}, \texttt{C12 B4} spells C-sharp to B, an ascending minor seventh. \textbf{Multi-step inference (n)} chains relations, for instance reinterpreting a chord enharmonically and then explaining its resolution and voice leading. These labels describe intended solution requirements, not calibrated difficulty. Table~\ref{tab:example-item} shows a complete level-n item.

Audio Q\&A crosses scope with difficulty. Every recording contributes one question pair in each of five configurations: global/easy, global/hard, local/easy, local/medium, and local/hard, giving 460 Chinese--English question pairs with 92 per configuration. A global/hard question may ask for the vocal range and where its extremes occur. A question may touch several musical aspects, and these labels are defined independently of the ABC solution hierarchy.

\begin{table}[!t]
\centering
\caption{ABC Q\&A item ABC-V2-0128 (global information, level n). Both language versions share the notation.}
\label{tab:example-item}
\begingroup\fontsize{9}{10.5}\selectfont
\XeTeXlinebreaklocale "zh"\XeTeXlinebreakskip=0pt plus 1pt
\setlength{\tabcolsep}{3pt}
\begin{tabular}{@{}p{0.13\columnwidth}p{0.84\columnwidth}@{}}
\toprule
Notation & \texttt{L:1/16 M:4/4 "A" z3 A4 A2F2 G4 D- |}\newline\texttt{"Dm" D3 D4 z2 D2 F4 G- |}\newline\texttt{"C6" G4- G A6 D- "Bb" D4- |}\\
Q (ZH) & {\cjkfont 仅凭三小节的和弦解决与后续色彩，判断被隐藏的主音和调式，并说明依据。}\\
Q (EN) & Using only the three-measure chord resolution and the color that follows, identify the hidden tonic and mode and explain the evidence.\\
Ref. & K:Dm, D minor. The A major triad is V in D minor and resolves to i, Dm; C6 and Bb supply the natural-minor colors $\flat$VII and VI, respectively, organizing the music around D rather than F.\\
\bottomrule
\end{tabular}
\endgroup
\end{table}

\subsection{Question Generation and Verification}

GPT-5.6 Sol with the \texttt{ultra} reasoning setting drafts ABC questions and references from the prepared materials; audio questions are instantiated from human-written templates (e.g., the chord progression of the verse) and take their references from the annotations. A reference gives the conclusion plus any requested explanation or rewritten notation. Reviewers check wording, musical correctness, and agreement among question, evidence, and reference, revising any of them before an item is accepted.

English versions of the accepted Chinese ABC and Audio questions and references are created over the same underlying evidence. The final ABC set holds 1,029 question pairs from 997 songs and 1,166 excerpts: 335 extraction, 441 single-step, and 253 multi-step items.

\begin{table*}[!t]
\centering
\caption{Audio Q\&A scores (0--4) on 460 question pairs from 92 recordings. Overall ZH/EN scores are separate; subgroup scores aggregate both languages. FCR is the percentage graded 4; bold marks the best per column.}
\label{tab:audio-results}
\begingroup
\fontsize{9}{10.5}\selectfont
\setlength{\tabcolsep}{2pt}
\renewcommand{\arraystretch}{1}

\begin{tabular*}{\textwidth}{@{\extracolsep{\fill}}lrrrrrrrrrrrr@{}}
\toprule
Model & \multicolumn{2}{c}{Overall score} & \multicolumn{4}{c}{Global} & \multicolumn{6}{c}{Local}\\
\cmidrule(lr){2-3}\cmidrule(lr){4-7}\cmidrule(l){8-13}
 & ZH & EN
 & \multicolumn{2}{c}{Easy (92)}
 & \multicolumn{2}{c}{Hard (92)}
 & \multicolumn{2}{c}{Easy (92)}
 & \multicolumn{2}{c}{Medium (92)}
 & \multicolumn{2}{c}{Hard (92)}\\
\cmidrule(lr){2-3}\cmidrule(lr){4-5}\cmidrule(lr){6-7}
\cmidrule(lr){8-9}\cmidrule(lr){10-11}\cmidrule(l){12-13}
 & & &
 Score & FCR &
 Score & FCR &
 Score & FCR &
 Score & FCR &
 Score & FCR\\
\midrule

Qwen 3.5 Omni
& \textbf{1.330} & \textbf{1.333}
& \textbf{2.522} & \textbf{48.37}
& \textbf{1.212} & \textbf{9.24}
& \textbf{1.815} & \textbf{23.37}
& 0.848 & 5.98
& \textbf{0.261} & 0.00\\

Gemini 3.1 Pro
& 1.176 & 1.122
& 2.250 & 44.02
& 0.924 & 4.35
& 1.554 & 22.28
& 0.891 & 10.33
& 0.125 & \textbf{0.54}\\

Gemini 3.7 Flash
& 1.159 & 1.028
& 1.859 & 35.87
& 0.913 & 4.89
& 1.543 & 22.83
& \textbf{1.000} & \textbf{10.87}
& 0.152 & \textbf{0.54}\\

Seed Lite
& 1.093 & 0.941
& 1.891 & 35.33
& 0.815 & 2.72
& 1.511 & 20.11
& 0.766 & 7.61
& 0.103 & 0.00\\

Qwen 3 Omni
& 0.965 & 0.913
& 1.630 & 26.63
& 0.929 & 3.80
& 1.429 & 20.11
& 0.636 & 4.35
& 0.071 & 0.00\\

StepAudio 3
& 0.946 & 0.863
& 1.647 & 27.72
& 0.799 & 2.17
& 1.375 & 19.57
& 0.641 & 7.07
& 0.060 & 0.00\\

Gemini 2.5 Pro
& 0.917 & 0.898
& 1.755 & 30.43
& 0.826 & 2.17
& 1.217 & 14.67
& 0.674 & 3.80
& 0.065 & 0.00\\

\bottomrule
\end{tabular*}

\par\vspace{3pt}
\begin{minipage}{\textwidth}
\fontsize{9}{10.5}\selectfont
Each configuration contains 92 question pairs, one per recording. Model names use the abbreviations in Table~\ref{tab:abc-results}.
\end{minipage}
\endgroup
\end{table*}

\section{Evaluation Protocol}
\label{sec:usage}

\subsection{Inputs and Responses}

Figure~\ref{fig:usage} summarizes evaluation. An ABC query consists of a question in Chinese or English and its notation excerpts; the paired versions share the notation, so language comparisons rest on the same evidence. The model returns an open response, including an explanation or rewritten ABC when asked. References and grading criteria are never shown to the answering model.

An audio query consists of the complete recording and a question in Chinese or English; the paired versions share the same recording and localization information. Local questions additionally name the target section, timestamp window, and bar range, while the full recording remains available as context. Human annotations and adjusted ABC support construction and scoring but are not provided to the answering model.

\subsection{Scoring and Metrics}

Qwen 3.5 Flash grades each response using the question and reference. For ABC Q\&A it also sees the notation; for Audio Q\&A it sees a structured answer key and textual annotation evidence but not the recording. Audio keys list required facts, accepted alternatives, and disallowed confusions.

Grades are integers: \textbf{0}, incorrect, unsupported, or missing; \textbf{1}, mostly incorrect with limited relevant content; \textbf{2}, partially correct with an important error or omission; \textbf{3}, correct core conclusion with minor omissions or imprecision; \textbf{4}, complete and correct. Musically equivalent expressions are accepted; missing a required explanation or component prevents full credit. After automatic grading, reviewers audited the assigned scores and evaluator rationales. About 20 retained items with ambiguous references or incomplete grading evidence were revised and re-evaluated, and all reported scores use the revised references and grading evidence.

For $N$ questions with grades $s_i$, we report the mean $\bar{s}=\frac{1}{N}\sum_i s_i$ and, for audio, the full-credit rate $\mathrm{FCR}=\frac{100}{N}\sum_i \mathbb{I}[s_i{=}4]$. Overall scores weight questions equally and are reported separately for Chinese and English. All subgroup scores aggregate both languages. The five audio configurations comprise 184 global and 276 local question pairs.

\section{Experiments}
\label{sec:evaluation}

Both parts evaluate seven omni models: Gemini 3.7 Flash~\cite{gemini37flash}, Gemini 3.1 Pro~\cite{gemini31pro}, Gemini 2.5 Pro~\cite{gemini25pro}, Seed2.0 Lite~\cite{seed20}, Qwen 3.5 Omni Plus~\cite{qwen35omni}, Qwen 3 Omni Flash~\cite{qwen3omni}, and StepAudio 3 Realtime~\cite{stepaudio3}. GPT-6 Astra~\cite{gpt6astra} and DeepSeek V4.1 Flash~\cite{deepseekv41flash} serve as text-only reference systems on ABC Q\&A. In total we grade 18,522 ABC responses (1,029 question pairs, two languages, nine models) and 6,440 audio responses (460 question pairs, two languages, seven models), all with the LLM evaluator of Section~\ref{sec:usage} and the final references.

\subsection{Symbolic Understanding: ABC Q\&A}

\textbf{Overall.} Table~\ref{tab:abc-results} gives overall scores by language and breakdowns by domain and solution requirement. Gemini 3.7 Flash leads the omni models in both languages (3.840/3.864), followed by Seed Lite (3.516/3.521) and StepAudio 3 (3.306/3.408); the text-only GPT-6 Astra reaches 3.964/3.945.

\textbf{Domains.} Table~\ref{tab:abc-results} shows the domain breakdowns. Among the omni models, Gemini 3.7 Flash has the highest mean in every domain, and every omni model scores highest on global information. Pitch and melody is the weakest domain for six of the seven omni models; Qwen 3 Omni Flash is instead weakest on verifiable symbolic tasks. Domain means also reflect question content and sample size, not domain difficulty in general.

\textbf{Language.} Neither language is systematically easier. Qwen 3.5 Omni Plus scores 0.119 lower in English than in Chinese, while StepAudio 3 scores 0.102 higher; the remaining omni models show smaller mixed differences. These results are descriptive language-wise comparisons and do not isolate translation or evaluator effects.

\textbf{Solution level.} The intended hierarchy does not order observed scores: Seed Lite scores 3.560 on extraction and 3.482 on multi-step inference, whereas Gemini 3.7 Flash scores 3.807 and 3.893 and Gemini 3.1 Pro 3.113 and 3.441. Content is confounded with level, since 165 of the 253 multi-step questions concern tonality and harmony and only five concern pitch and melody, so level means are not a controlled comparison of reasoning depth.

\textbf{Grading symbolic output.} ABC-V2-0002 asks for a major-second transposition down from C-sharp minor to B minor; a correct answer must transpose pitches and chord symbols while preserving durations, rests, and ties. Seed Lite, Qwen 3.5 Omni Plus, and StepAudio 3 each receive 4 for responses in which one to three reference notes written as \texttt{B,} are rendered as \texttt{B}, one octave too high. These examples illustrate why high LLM-assigned scores on symbolic output need event-level checks; they do not estimate the evaluator's overall error rate.

\subsection{Audio Understanding: Audio Q\&A}

\textbf{Overall.} Table~\ref{tab:audio-results} gives overall mean scores by language and scope--difficulty results per configuration. Qwen 3.5 Omni Plus has the highest observed overall score in both Chinese and English (1.330/1.333), followed by Gemini 3.1 Pro (1.176/1.122) and Gemini 3.7 Flash (1.159/1.028). The observed bilingual audio and ABC rankings have Spearman $\rho=-0.04$ across the seven omni models; Qwen 3.5 Omni Plus, sixth of seven on ABC Q\&A, is first on Audio Q\&A. The two parts use different songs and tasks, so this is not a controlled modality comparison.

\textbf{Scope and difficulty.} Across the five configurations, Qwen 3.5 Omni Plus has the highest mean in four, while Gemini 3.7 Flash leads local/medium (1.000). Local/hard is the weakest configuration for every model (0.060--0.261). Every model also scores higher on global than on local questions (1.867 vs.\ 0.975 for Qwen 3.5 Omni Plus). Scope and difficulty are not varied independently of content, so these gaps are not isolated effects of localization or reasoning depth.

\textbf{Case study: harmonic function in context.} On the Chinese version, JAZZ16-V4-T02-Q03 asks which chord follows C7 at the end of the first chorus and how the resolution functions in the main key. The reference (locator 00:44.43--01:06.43) identifies C7--Fm as V7--i in F minor. Qwen 3.5 Omni Plus names both chord and function and receives 4. Gemini 3.7 Flash substitutes F/Fmaj7 and F major and receives 1. Gemini 2.5 Pro names Fm but places it in A-flat major as a deceptive V/V--vi resolution, also receiving 1: naming a chord and grounding its function in the tonal context are distinct requirements.

\subsection{Limitations}

The benchmark uses only Suno-generated music and lacks question-only controls~\cite{rulistening}. In addition, 82 audio references await further listening review, human--grader agreement has not been measured, and affected ABC models require re-evaluation under a single logged prompt. Completing these checks and extending evaluation to broader music sources are priorities for future work.

\section{Conclusion}
\label{sec:conclusion}

HMU-Bench provides bilingual, open-ended evaluation of notation and audio understanding, with explicit labels for solution requirements. Across seven omni models, the strongest systems score highly on ABC Q\&A, while audio scores remain low and model rankings differ substantially. Because the two parts use different songs and tasks, this gap is not a controlled comparison of modalities. Cases where LLM grading misses octave errors further highlight the need for event-level validation of symbolic outputs alongside reference-based evaluation.

\clearpage
\nobalance
\section{Acknowledgment}
AI tools were used to polish the manuscript's language. This editing did not contribute to the technical content of the work.


\section{Compliance with Ethical Standards}
This study did not require ethical approval under our institutional guidelines.

\bibliographystyle{IEEEbib}
\bibliography{references}
\end{document}